\documentclass[a4paper,oneside,11pt]{article}

\usepackage{amsfonts,color}

\newtheorem{dfn}{Definition}[section]
\newtheorem{tw}[dfn]{Theorem}
\newtheorem{prop}[dfn]{Proposition}

\newtheorem{lem}[dfn]{Lemma}

\usepackage{amssymb} 
\usepackage{amsmath}
\numberwithin{equation}{section}

\usepackage{anysize}
\usepackage{array}
\usepackage{enumerate}
\author{Micha\l \ Barski \\ \small  Faculty of Mathematics and Computer Science, University of Leipzig, Germany\\
\small Faculty of Mathematics, Cardinal Stefan Wyszy\'nski University in Warsaw, Poland
\\ \small{\it Michal.Barski@math.uni-leipzig.de}}

\title{\bf Incompleteness of the bond market with L\'evy noise under the physical measure}

\begin{document}


\baselineskip=16pt
\maketitle
\date

\begin{abstract}
The problem of completeness of the forward rate based bond market model driven by a L\'evy process under the physical measure is examined. The
incompleteness of  market in the case when the L\'evy measure has a density function is shown. The required elements of the theory of stochastic integration over the compensated jump measure under a martingale measure is presented and the corresponding integral representation of local martingales is proven.
\end{abstract}

\noindent
\begin{quote}


\end{quote}


\section{Introduction} 
A bond with maturity $T\geq 0$ is a financial contract paying to its owner $1$ at the date $T$. The price of the bond $P(t,T), t\in[0,T]$ is a stochastic process satisfying $P(T,T)=1$ and the family $P(\cdot,T); \ T\in[0,T^\ast]$ forms a bond market with a finite time horizon $T^\ast<+\infty$. One possible approach to construct the bond market model is based on the random field $f(t,T); \ t,T\in[0,T^\ast]$ called forward rate. The prices are then defined by the exponential formula
\begin{gather}\label{P wstep}
P(t,T)=e^{-\int_{t}^{T}f(t,u)du}, \quad t\in[0,T], \ T\in[0,T^\ast].
\end{gather}
Random behaviour in the model is enforced by a L\'evy process $Z$ defined on a probability space $(\Omega,\mathcal{F},P)$ with filtration $(\mathcal{F}_t), t\in[0,T^\ast]$. For any $T\in[0,T^\ast]$ the forward rate process $f(\cdot,T)$ is defined by the dynamics of the form
\begin{gather}\label{df wstep}
df(t,T)=\alpha(t,T)dt+\sigma(t,T)dZ(t), \quad t\in[0,T].
\end{gather}
The measure $P$ under which the model is constructed will be called a {\it physical measure}. If $Z$ is a Wiener process then \eqref{P wstep}-\eqref{df wstep} provide the earliest form of the model introduced by Heath, Jarrow and Morton in \cite{HeatJarrMor} which was afterwards extensively studied in the literature. The modification involving a L\'evy process reflects better the real behaviour of the bond prices like, for instance, their heavy-tailed distributions. On the other side it also leads to new problems concerned with the definition of bond portfolios, option pricing and hedging which were absent in the no-jump setting. 

Let $X$ be an $\mathcal{F}_{T^\ast}$-measurable random variable which represents the payoff at time $T^{\ast}$ of a financial contract. A bond portfolio $\varphi$, which is to be precisely defined, {\it replicates} $X$ if the corresponding wealth process $X^{\varphi}$ satisfies
\begin{gather}\label{portfel replikujacy}
X^{\varphi}_{T^\ast}=X, \quad P-a.s.. 
\end{gather}
If each bounded payoff can be replicated then the market is called {\it complete} and {\it incomplete} in the opposite case.
The analysis of the problem \eqref{portfel replikujacy}, that is the issue of existence of $\varphi$, requires the passage to  the risk-neutral setting governed by the family of the so called martingale measures. Recall, $Q$ is a {\it martingale measure} if it is equivalent to $P$ and the discounted bond prices are $Q$-local martingales. Application of the Girsanov theorem, see \cite{Kun}, yields the dynamics of the forward rate under $Q$, which is
\begin{gather}\label{df przy Q wstep}
df(t,T)=\tilde{\alpha}(t,T)dt+\sigma(t,T)d\tilde{Z}(t), \quad t,T\in[0,T^\ast],
\end{gather}
where $\tilde{\alpha}(\cdot,\cdot)$ is a modified drift and $\tilde{Z}$ stands for the transformation of $Z$ under $Q$. 
If $Z=W$ is a Wiener process under $P$ so is $\tilde{Z}=\tilde{W}$ under $Q$ and the martingale representation theorem provides the integral decomposition
\begin{gather*}
M_t=M_0+\int_{0}^{t}\phi_M(s)d\tilde{W}(s), \quad t\in[0,T^\ast],
\end{gather*}
of the martingale $M_t=E^Q[X\mid\mathcal{F}_t], t\in[0,T^\ast]$. Above $\phi_M$ is a certain process and it enables to determine $\varphi$ which solves \eqref{portfel replikujacy}. If $Z$ is a general L\'evy process then the arguments above fail for two reasons. The first is that L\'evy processes are not stable under a measure change, that is $\tilde{Z}$ is no longer a L\'evy process under $Q$. Its increments may be not stationary nor independent.
Consequently, the forward rate dynamics \eqref{df przy Q wstep} has a non-L\'evy structure. The second reason, which in fact arises from the first one,
is that we need a relevant version of the martingale representation theorem under $Q$. A model framework which is commonly used in the literature and allows to overcome these two difficulties is to assume that $P$ is simultaneously a martingale measure. Then $Z=\tilde{Z}$ and any local martingale can be represented in the form
\begin{gather}\label{rozklad martyngalu wstep}
M_t=M_0+\int_{0}^{t}\phi_M(s)dW(s)+\int_{0}^{t}\int_{\mathbb{R}}\psi_M(s,y)\tilde{\pi}(ds,dy), \quad t\in[0,T^\ast].
\end{gather}
with some $\phi_M, \psi_M$, see \cite{Kun}. Above $\tilde{\pi}$ stands for the compensated jump measure of $Z$ under 
$P$. As was shown in \cite{BarZab IJTAF} the existence of $\phi_M, \psi_M$ for $M_t:=E[X\mid\mathcal{F}_t]$ does not imply the existence of $\varphi$ solving \eqref{portfel replikujacy}, that is there exists a financial contract $X$ which can not be replicated. The problem \eqref{portfel replikujacy}, in the case when the physical measure $P$ is not a martingale measure, has not been examined in the literature. 

In this paper we investigate the problem \eqref{portfel replikujacy} without the assumption that $P$ is a martingale measure. We provide a systematic treatment of the issue of passage from the physical measure $P$ to the martingale one $Q$ and prove a required version of the martingale representation theorem which allows to write any $Q$-local martingale $M$ in the form
\begin{gather}\label{rozklad martyngalu przy Q wstep}
M_t=M_0+\int_{0}^{t}\phi_M(s)d\tilde{W}(s)+\int_{0}^{t}\int_{\mathbb{R}}\psi_M(s,y)\tilde{\pi}_Q(ds,dy), \quad t\in[0,T^\ast],
\end{gather}
where $\tilde{\pi}_Q$ is a compensated jump measure of $Z$ under $Q$. In particular we present precisely the construction of the second integral in \eqref{rozklad martyngalu przy Q wstep}. Our main result is Theorem \ref{tw_o_niezupelnosci} in Section \ref{Incompleteness} showing that there exists a bounded random variable $X$ for which \eqref{portfel replikujacy} has no solution providing that the L\'evy measure of $Z$ has a density function. This means that then the bond market model is incomplete, no matter  if the martingale measure is unique or not. The result implies that in the bond market the classical relation known from stock markets between the uniqueness of the martingale measure and completeness brakes down.

The paper consists of three parts. In Section \ref{Levy process and related martingale representation} we discuss properties of a L\'evy process which are needed to formulate the martingale decomposition formula \eqref{rozklad martyngalu wstep} and further to describe equivalent measures. The construction of a stochastic integral over the compensated jump measure under an equivalent measure and the related martingale representation formula \eqref{rozklad martyngalu przy Q wstep} are presented in Section \ref{Martingale representation under equivalent measures}. The incompleteness of the bond market is treated in Section \ref{Incompleteness of the bond market} where we precisely introduce the bond market model, the concept of a bond portfolio and finally prove Theorem \ref{tw_o_niezupelnosci}.  
 
\section{L\'evy process and related martingale representation}\label{Levy process and related martingale representation}
We start with summarizing properties of L\'evy processes which are needed in the paper. Their proofs can be found, for instance, in \cite{Appl}. 

Let $Z$ be a real valued L\'evy process on a probability space $(\Omega,\mathcal{F}, P)$ with filtration $(\mathcal{F}_t), t\in[0,T^\ast]$ such that $\mathcal{F}_{T^\ast}=\mathcal{F}$. It is known that $Z$ has a modification with c\`adl\`ag trajectories and only this modification will be considered in the sequel. For any $\varepsilon>0$ the number of jumps on $[0,T^\ast]$ such that $\mid\triangle Z_s\mid:=\mid Z_s-Z_{s-}\mid>\varepsilon$ is finite almost surely. Consequently, for any $A\subseteq\mathbb{R}$ which is separated from zero, that is $0\notin\bar{A}$, where $\bar{A}$ stands for the closure of $A$, the random variable 
$$
\pi(t,A):=\sharp\{s\in[0,t]: \triangle Z_s\in A\}, \quad t\in[0,T^\ast],
$$
is well defined. It counts the number of jumps of $Z$ on the interval $[0,t]$ which lie in the set $A$. The function $\pi(\cdot,\cdot)$ can be treated as a $\sigma$-finite measure on $[0,T^\ast]\times \mathbb{R}$. It is called a {\it jump measure} of $Z$.  From the independence and stationarity of the increments of $Z$ follow two important properties of the jump measure, that is for any $A,B$ separated from zero hold
\begin{align}\label{poissonowskosc miary skokow 1}
\pi(t,A), t\in[0,T^\ast] \ \text{is a Poisson process with intensity} \ \lambda_A:=E[\pi(1,A)],\\[1ex]\label{poissonowskosc miary skokow 2}
\text{For any} \ t\in[0,T^\ast] \ \text{the r.v.} \  \pi(t,A),\pi(t,B) \ \text{are independent if} \  A\cap B=\emptyset.
\end{align}
The $\sigma$-finite measure $\nu$ on $\mathbb{R}$ defined by 
\begin{gather}\label{nu definicja}
\nu(A):=E[\pi(1,A)], \quad 0\notin\bar{A},
\end{gather} 
is called an {\it intensity measure} or a {\it L\'evy measure} of $Z$. It satisfies the integrability condition 
\begin{gather}\label{warunek na miare Levyego}
\int_{\mathbb{R}}(\mid y\mid^2\wedge \ 1)\nu(dy)<+\infty.
\end{gather}
Because of \eqref{poissonowskosc miary skokow 1}-\eqref{poissonowskosc miary skokow 2} and \eqref{nu definicja} the measure $\pi$ is called a {\it Poisson random measure} with intensity measure $\nu$. On the other hand any measure satisfying \eqref{warunek na miare Levyego} is an intensity measure of some Poisson random measure. Further, it follows that, for a separated from zero set $A$, the process
$$
\tilde{\pi}(t,A):=\pi(t,A)-t\nu(A),  \quad t\in[0,T^\ast],
$$
is a martingale, which means that $dt\nu(dy)$ is a {\it compensating measure} for $\pi(dt,dy)$. The measure $\tilde{\pi}(dt,dy)$ is called a {\it compensated jump measure} of $Z$. For $f:\mathbb{R}\longrightarrow\mathbb{R}$, a set $A$ separated from zero and any $t\in[0,T^\ast]$ the random variable
$$
\int_{0}^{t}\int_{A}f(y)\pi(ds,dy)=\sum_{s\in[0,t]}f(\triangle Z_s)\mathbf{1}_{A}(\triangle Z_s),
$$ 
is integrable with expectation
$$
E\Big(\int_{0}^{t}\int_{A}f(y)\pi(ds,dy)\Big)=t\int_{A}f(y)\nu(dy).
$$
Further, the process $\int_{0}^{t}\int_{A}f(y)\tilde{\pi}(ds,dy)$ is a square integrable martingale and 
\begin{gather}\label{drugi moment calki skompensowanej}
E\Big(\Big(\int_{0}^{t}\int_{A}f(y)\pi(ds,dy)\Big)^2\Big)=t\int_{A}f^2(y)\nu(dy),\quad t\in[0,T^\ast].
\end{gather}
For $f(y)=y$ and a sequence of sets $A_n:=\{\frac{1}{n}<\mid y\mid\leq 1\}$ one can prove, using
\eqref{drugi moment calki skompensowanej} and \eqref{warunek na miare Levyego}, that the sequence 
$$
\int_{0}^{t}\int_{A_n}y \ \tilde{\pi}(ds,dy),  \quad t\in[0,T^\ast], n=1,2,...,
$$ 
converges almost surely uniformly on $[0,T^\ast]$. The limit is denoted by
$$
\int_{0}^{t}\int_{\{\mid y\mid\leq1\}}y \ \tilde{\pi}(ds,dy):=\lim_{n\rightarrow +\infty}\int_{0}^{t}\int_{A_n}y \ \tilde{\pi}(ds,dy), \quad t\in[0,T^\ast].
$$
Now we are ready to formulate the L\'evy-It\^o decomposition. It tells that any  L\'evy process $Z$ admits the following representation
\begin{gather}\label{tw Levy Ito decomposition}
Z_t=at+W(t)+\int_{0}^{t}\int_{\{\mid y\mid\leq 1\}}y \ \tilde{\pi}(ds,dy)+\int_{0}^{t}\int_{\{\mid y\mid>1\}}y \ \pi(ds,dy), \quad t\in[0,T^\ast],
\end{gather}
where $a\in \mathbb{R}$, $W$ is a Wiener process with variance $q>0$, that is $Var (W_t)=qt$. Moreover, all the ingredients in 
\eqref{tw Levy Ito decomposition} are independent. 
The L\'evy-It\^o decomposition is an important tool in the analysis of L\'evy processes. One of its consequences is that it makes possible 
to define the stochastic integral 
\begin{gather}\label{calka po dZ}
\int_{0}^{t}f(s)dZ(s), \quad t\in[0,T^\ast],
\end{gather}
for integrands $f\in\Phi$, where $\Phi$ is a family of predictable and square integrable processes, i.e. such that
$$
\int_{0}^{T^\ast}\mid f(s)\mid^2ds<+\infty.
$$
 The definition of the class $\Phi$ is commonly known if $Z$ is a Wiener process. The passage to the general case is based on the fact that the first integral on the right side of \eqref{tw Levy Ito decomposition} is a square integrable martingale. In the case when $Z$ is a martingale, that is when 
 $$
 \int_{\{\mid y\mid>1\}}\mid y\mid \nu(dy)<+\infty, \quad \text{and} \quad  a=- \int_{\{\mid y\mid>1\}} y \ \nu(dy),
 $$
the form of $Z$ is
$$
Z_t=W(t)+\int_{0}^{t}\int_{\mathbb{R}}y \ \tilde{\pi}(ds,dy), \quad t\in[0,T^\ast],
$$
and then the integral  \eqref{calka po dZ} is a local martingale. It turns out that the class $\Phi$ is to narrow to represent any local martingale as a stochastic integral \eqref{calka po dZ} with some $f\in\Phi$. However, the integral representation of local martingales is possible in the class of integrals 
$$
\int_{0}^{t}f(s)dW(s), \quad \int_{0}^{t}\int_{\mathbb{R}}g(s,y)\tilde{\pi}(ds,dy).
$$
In the following section we present the construction of the second integral above. Afterwards, in Section \ref{section Martingale representation and characterisation of equivalent measures} we formulate the representation theorem for local martingales.

\subsection{Integration over the compensated jump measure}\label{section Construction of the integral}
Here we present the construction of the integral
$$
\int_{0}^{t}\int_{U}g(s,y) \ \tilde{\pi}(ds,dy), \quad t\in[0,T^\ast],
$$
where $g:[0,T^\ast]\times \mathbb{R}\rightarrow\mathbb{R}$ and $\tilde{\pi}$ stands for a compensated jump measure of the L\'evy process $Z$. We start from an intuitive definition of the integral for simple integrands and further extend it for integrands satisfying certain integrability conditions. The procedure provides a class of integrands which allow to obtain the integral representation for any local martingale which is discussed in Section \ref{section Martingale representation and characterisation of equivalent measures}. The construction presented below has been sketched in \cite{Kun}.  Our presentation contains more details since 
it will serve as a point of reference for the extension of the concept of integration under an equivalent measure in Section \ref{section Construction of the integral}.

The process $g=g(t,y)$ is {\it simple} if it has the form 
\begin{gather}\label{g simple}
 g(s,y)=g(0,y)\mathbf{1}_{\{s=0\}}+\sum_{i=0}^{n-1}\left(\sum_{j=1}^{m_i}g_{ij}\mathbf{1}_{(t_i,t_{i+1}]}(s)\mathbf{1}_{A_{ij}}\right), \qquad s\in[0,T^\ast], \ y\in \mathbb{R},
\end{gather}
where $0=t_0<t_1<...<t_n=T^\ast$ is a partition of $[0,T^\ast]$ and $A_{ij}$ is a family of sets of $\mathbb{R}$ which are separated from zero, i.e.
$$
0\notin \bar{A}_{ij}. 
$$
For a given subinterval $(t_i,t_{i+1}]$ the process $g$ is a linear combination of the terms $g_{ij}\mathbf{1}_{(t_i,t_{i+1}]}(s)\mathbf{1}_{A_{ij}}$, where $g_{ij}$ are bounded $\mathcal{F}_{t_i}$- measurable random variables and $A_{ij}, j=1,2,...,m_i$ are disjoint. Notice that we do not assume that the sets $A_{ij}$ and $A_{kl}$ are disjoint for $i\neq k$. Denote the class of all simple processes by $\mathcal{S}$. For $g\in\mathcal{S}$ a stochastic integral $I(g)$ is defined by
$$
I(g)_t=\int_{0}^{t}\int_{U}g(s,y)\tilde{\pi}(ds,dy):=\sum_{i=0}^{n}\sum_{j=1}^{m_i}g_{ij}\tilde{\pi}((t_i\wedge t,t_{i+1}\wedge t]\times A_{ij}), \qquad t\in[0,T^\ast].
$$
We will show that $I(g)$ is a square integrable martingale and find its second moment. It follows from \eqref{poissonowskosc miary skokow 1}-\eqref{poissonowskosc miary skokow 2} that the processes
$$
\tilde{\pi}(t,A_{ij})=\pi(t,A_{ij})-t\nu(A_{ij}), \quad t\in[0,T^\ast],
$$
are square integrable martingales with independent increments and $\tilde{\pi}(t,A_{ij})$, $\tilde{\pi}(t,A_{kl})$ are independent if $A_{ij}\cap A_{kl}=\emptyset$.  As a direct consequence of that we obtain the following result.
\begin{prop}\label{prop o wwo under P}
 For the sets $A,B\subseteq \mathbb{R}$ separated from zero and $s<t, s,t\in[0,T^\ast]$ hold  
 \begin{align*}
  &E[\tilde{\pi}^2\left((s,t]\times A\right)\mid\mathcal{F}_{s}]=(t-s)\nu(A),\\[1ex]
  &E[\tilde{\pi}\left((s,t]\times A\right)\cdot \tilde{\pi}\left((s,t]\times B\right)\mid \mathcal{F}_s]=0, \quad \text{if} \ A\cap B=\emptyset, \\[1ex]
  &E[\tilde{\pi}\left((s,t]\times A\right)\cdot \tilde{\pi}\left((u,v]\times B\right)\mid \mathcal{F}_u]=0, \quad \text{for} \ t\leq u<v\leq T^\ast.
 \end{align*}
\end{prop}
Proposition \ref{prop o wwo under P} is a key toll for proving the isometric formula below.    
\begin{prop}\label{prop izometria dla prostych przy P}
 For $g\in\mathcal{S}$ the integral $I(g)$ is a square integrable martingale and 
 \begin{gather}\label{wzor izometryczny dla prostych przy P}
   E\left[\mid I(g)_t\mid^2\right]=E\left[\int_{0}^{t}\int_{\mathbb{R}}\mid g(s,y)\mid^2ds\nu(dy))\right], \quad t\in[0,T^\ast].
 \end{gather}
\end{prop}
{\bf Proof:} It is clear that $I(g)$ is a martingale. For the sake of simplicity we prove \eqref{wzor izometryczny dla prostych przy P} for $t=T^\ast$ only. 
From the definition of the integral follows
$$
E\Big[\mid I(g)_{T^\ast}\mid^2]=E\Big[\sum_{i=0}^{n}\sum_{j=1}^{m_i}\sum_{k=0}^{n}\sum_{l=1}^{m_l} g_{ij}g_{kl}\cdot\tilde{\pi}((t_i,t_{i+1}]\times A_{ij})\cdot \tilde{\pi}((t_k,t_{k+1}]\times A_{kl})\Big].
$$
Using Proposition \ref{prop o wwo under P} let us calculate the expectations of the terms appearing in the above sum.
We need to consider the following three cases:
\begin{enumerate}[a)]
 \item if $i=k$ and $j=l$ then
 \begin{align*}
 E[g_{ij}g_{ij}\cdot\tilde{\pi}^2((t_i,t_{i+1}]\times A_{ij})]&=E\Big[\mid g_{ij}\mid^2 E[\tilde{\pi}^2((t_i,t_{i+1}]\times A_{ij})\mid\mathcal{F}_{t_i}]\Big]\\[1ex]
 &=E\big[\mid g_{ij}\mid^2\ (t_{i+1}-t_i)\nu(A_{ij})\big],
 \end{align*}
 \item if $i=k$ and $j\neq l$ then
 \begin{align*}
 &E[g_{ij}g_{il}\cdot\tilde{\pi}((t_i,t_{i+1}]\times A_{ij})\cdot \tilde{\pi}((t_i,t_{i+1}]\times A_{il})]\\[1ex]
 &=E\Big[g_{ij}g_{il} \cdot E[\tilde{\pi}((t_i,t_{i+1}]\times A_{ij})\tilde{\pi}((t_i,t_{i+1}]\times A_{il})\mid\mathcal{F}_{t_i}]\Big]=0,
  \end{align*}
  \item if $i\neq k$ then
 \begin{align*}
 &E[g_{ij}g_{kl}\cdot\tilde{\pi}((t_i,t_{i+1}]\times A_{ij})\cdot \tilde{\pi}((t_k,t_{k+1}]\times A_{kl})]\\[1ex]
 &=E\Big[g_{ij}g_{kl}\cdot E[\tilde{\pi}((t_i,t_{i+1}]\times A_{ij})\tilde{\pi}((t_k,t_{k+1}]\times A_{kl})\mid\mathcal{F}_{t_k\vee t_i}]\Big]=0.
  \end{align*}
\end{enumerate}
From the above follows 
\begin{align*}
E\Big[\big\vert\int_{0}^{T^\ast}\int_{\mathbb{R}}g(s,y)\tilde{\pi}(ds,dy)\big\vert^2\Big]&=\sum_{i=0}^n\sum_{j=1}^{m_i}E[\mid g_{ij}\mid^2\ (t_{i+1}-t_i)\nu(A_{ij})]\\[1ex]
&=E\Big[\int_{0}^{T^\ast}\int_{\mathbb{R}}\mid g(s,y)\mid^2ds\nu(dy)\Big],
\end{align*}
which is \eqref{wzor izometryczny dla prostych przy P}.
\hfill$\square$

\vskip2mm
The definition of the integral can be extended to the class of all predictable process satisfying
$$
E\Big[\int_{0}^{T^\ast}\int_{\mathbb{R}}\mid g(s,y)\mid^2 ds\nu(dy)\Big]<+\infty.
$$
If this is the case then there exists a sequence $g_n\in\mathcal{S}, n=1,2,...$ such that 
$$
E\Big[\int_{0}^{T^\ast}\int_{\mathbb{R}}\mid g(s,y)-g_n(s,y)\mid^2 ds\nu(dy)\Big]\longrightarrow 0,
$$
which implies that 
$$
E[\mid I(g_n)_{T^\ast}-I(g_m)_{T^\ast}\mid^2]\underset{n,m}{\longrightarrow} 0.
$$
The condition above tells that $\{I(g_n)\}$ is a Cauchy sequence in the space of square integrable martingales which is complete. Thus there exists a limit $I(g)$ and it defines the integral for the integrator $g$, that is
$$
\int_{0}^{t}g(s,y)\tilde{\pi}(ds,dy)=I(g)_t:=\lim_{n\rightarrow +\infty}I(g_n)_{t},
$$
and the isometric formula \eqref{wzor izometryczny dla prostych przy P} still holds.
Let us introduce a class $\Psi_2$ of all predictable processes satisfying
$$
\Psi_2: \qquad \int_{0}^{T^\ast}\int_{\mathbb{R}}\mid g(s,y)\mid^2 ds\nu(dy)<+\infty, \quad P-a.s..
$$
By using the localizing arguments one can show that for $g\in\Psi_2$ the integral
$$
\int_{0}^{t}\int_{\mathbb{R}}g(s,y)\tilde{\pi}(ds,dy), \quad t\in[0,T^\ast], 
$$
is a well defined locally square integrable martingale.

The second class of process which are $\tilde{\pi}$-integrable consists of all predictable ones such that 
\begin{gather}\label{EP skonczona}
E\left[\int_{0}^{T^\ast}\int_{\mathbb{R}}\mid g(s,y)\mid ds\nu(dy)\right]<+\infty.
\end{gather}
Then it follows from the definition of the compensating measure that 
\begin{gather}\label{def skompensowanej calki przy P}
\int_{0}^{t}\int_{\mathbb{R}}g(s,y)\tilde{\pi}(ds,dy):=\int_{0}^{t}\int_{\mathbb{R}}g(s,y){\pi}(dy,ds)-\int_{0}^{t}\int_{\mathbb{R}}g(s,y)ds\nu(dy), \qquad t\in[0,T^\ast],
\end{gather}
is a martingale. In the class $\Psi_1$ of predictable processes satisfying
$$
\Psi_1: \qquad \int_{0}^{T^\ast}\int_{\mathbb{R}}\mid g(s,y)\mid ds\nu(dy)<+\infty, \qquad P-a.s.,
$$
the condition \eqref{EP skonczona} holds locally and thus
$$
\int_{0}^{t}\int_{\mathbb{R}}g(s,y)\tilde{\pi}(ds,dy), \quad t\in[0,T^\ast],
$$
is a local martingale.
The class of processes which plays the crucial role in representing local martingales is $\Psi_{1,2}$ defined below
$$
g\in\Psi_{1,2} \qquad \Longleftrightarrow \qquad g\mathbf{1}_{\{\mid g\mid\leq 1\}}\in\Psi_2 \quad \text{and}\quad g\mathbf{1}_{\{\mid g\mid> 1\}}\in\Psi_1.
$$
For each $g\in\Psi_{1,2}$ the integral is defined by the decomposition
$$
\int_{0}^{t}\int_{\mathbb{R}}g(s,y)\tilde{\pi}(ds,dy):=\int_{0}^{t}\int_{\mathbb{R}}g\mathbf{1}_{\{\mid g\mid\leq 1\}}\tilde{\pi}(ds,dy)
+\int_{0}^{t}\int_{\mathbb{R}}g\mathbf{1}_{\{\mid g\mid> 1\}}\tilde{\pi}(ds,dy),
$$
and is a local martingale. The class $\Psi_{1,2}$ can be described alternatively by the condition
$$
\Psi_{1,2}: \qquad \int_{0}^{T^\ast}\int_{\mathbb{R}}(\mid g(s,y)\mid^2\wedge \mid g(s,y)\mid)ds\nu(dy)<+\infty, \quad P-a.s..
$$

\subsection{Martingale representation and characterisation of equivalent measures}\label{section Martingale representation and characterisation of equivalent measures}

Let $\Phi$ stand for the class of processes integrable with respect to the Wiener process, that is $\phi\in\Phi$
 if $\phi$ is predictable and satisfies
 $$
 \int_{0}^{T^\ast}\mid\phi(s)\mid^2ds<+\infty, P-a.s..
 $$
 For any $\phi\in\Phi$ the integral
 $$
 \int_{0}^{t}\phi(s)dW(s), \quad t\in[0,T^\ast],
 $$
 is a local square integrable martingale. The classes of integrands $\Phi$ and $\Psi_{1,2}$ introduced in Section \ref{section Construction of the integral} are sufficiently large to represent local martingales as stochastic integrals. The result below has been proven in \cite{Kun}.

\begin{tw}\label{tw Kunity}
Let $M$ be an $\mathbb{R}$-valued $P$-local martingale on $[0,T^\ast]$. Then there exist
$\phi_M\in\Phi$ and $\psi_M\in\Psi_{1,2}$ satisfying
\begin{gather}\label{reprezentacja Kunity}
M_t=M_0+\int_{0}^{t}\phi_M(s) dW(s)+\int_{0}^{t}\int_{U}\psi_M(s,y)\tilde{\pi}(ds,dy), \qquad t\in[0,T^\ast].
\end{gather}
Moreover, the pair $(\phi_M,\psi_M)$ is unique i.e., if
$(\phi_M^{\prime},\psi_M^{\prime})$ satisfies \eqref{reprezentacja Kunity} then
\begin{gather*}
\phi_M=\phi_M^{\prime}, \quad dP\times dt- \ \text{a.s.}  \quad
\text{and} \quad \psi_M=\psi_M^{\prime},   \quad
dP\times dt\times d\nu- \ \text{a.s.}.
\end{gather*}
\end{tw}

Let us consider  a measure on $(\Omega,\mathcal{F})$ which is equivalent to $P$. The equivalence implies the existence of a positive density process which can be written in the form
\begin{gather}\label{gestosc w postaci eksponencjalnej}
\rho_t:=\frac{dQ}{dP}\Big\arrowvert_{\mathcal{F}_t}=e^{Y_t}, \qquad t\in[0,T^\ast],
\end{gather}
with $Y$ such that $\rho$ is a martingale under $P$. The following generalization of the classical Girsanov theorem provides an explicit form of the integral representation of $\rho$ and characterizes the process $Z$ under the measure $Q$, for the proof see \cite{Kun}.

\begin{tw}[Girsanov]\label{tw Girsanova}
Let $Q\sim P$ and $Z$ be a L\'evy process under $P$ with a characteristic triplet $(a,q,\nu)$. 
\begin{enumerate}[a)]
\item There exists a pair of processes $(\phi,\psi)$ such that $\phi\in\Phi$ and $e^\psi-1\in\Psi_{1,2}$ such that the density process \eqref{gestosc w postaci eksponencjalnej} has the form
\begin{gather}\label{wzor na gestosc}
d\rho(t)=\rho(t-)\left[\phi(t) dW(t)+\int_{\mathbb{R}}(e^{\psi(t,y)}-1)\tilde{\pi}(dt,dy)\right], \quad \rho(0)=1, \quad t\in[0,T^\ast],
\end{gather}
with $E[\rho_t]=1, t\in[0,T^\ast]$. 
\item Under the measure $Q$ the process
\begin{gather}\label{tilde W}
\tilde{W}(t):=W(t)-\int_{0}^{t}\phi(s)ds, \quad t\in[0,T^\ast],
\end{gather}
is a Wiener process with variance $q$ and the random measure
\begin{gather}\label{nu_Q}
\nu_Q(dt,dy):=e^{\psi(t,y)}dt \nu(dy), \quad t\in[0,T^\ast], y\in \mathbb{R},
\end{gather}
is a compensating measure for the jump measure $\pi(dt,dy)$ of $Z$.
\item  Under the measure $Q$ the process $Z$ admits 
the representation
\begin{gather}\label{Z przy mierze Q}
Z(t)=\tilde{a}_t+\tilde{W}(t)+\int_{0}^{t}\int_{\{\mid y\mid\leq 1\}}y \ \tilde{\pi}_Q(ds,dy)+\int_{0}^{t}\int_{\{\mid y\mid> 1\}}y \ \pi(ds,dy),
\end{gather}
with 
$$
\tilde{a}_t:=at+\int_{0}^{t}\phi(s)ds+\int_{0}^{t}\int_{\{\mid y\mid\leq 1\}}y(e^{\psi(s,y)}-1)ds\nu(dy).
$$
\end{enumerate}
\end{tw}
\vskip2mm
\noindent
A pair $(\phi,\psi)$ appearing in the theorem will be called a {\it generating pair} of the measure $Q$. The Dol\'eans-Dade equation \eqref{wzor na gestosc} can be solved explicitly to see that the density process $\rho $ is actually like in \eqref{gestosc w postaci eksponencjalnej} with
\begin{align}\label{wzor Y}\nonumber
Y(t)&=\int_{0}^{t}\phi(s)dW(s)-\frac{1}{2}\int_{0}^{t}\mid \sqrt{q}\phi(s)\mid^2ds\\[2ex]
&+\int_{0}^{t}\int_{\mathbb{R}}(e^{\psi(s,y)}-1)\tilde{\pi}(ds,dy)-\int_{0}^{t}\int_{\mathbb{R}}(e^{\psi(s,y)}-1-\psi(s,y))\pi(ds,dy).
\end{align}
To comment on the theorem let us introduce 
two sets $\underline{A}=\underline{A}(t)$ and $\hat{A}=\hat{A}(t)$ by
 $$
\underline{A}:=\{(s,y)\in[0,t]\times \mathbb{R}: \mid e^{\psi(s,y)}-1\mid\leq 1\}, \quad
\hat{A}:=\{(s,y)\in[0,t]\times \mathbb{R}: \mid e^{\psi(s,y)}-1\mid> 1\}.
$$
First, let us explain that \eqref{Z przy mierze Q} is well defined. Indeed, from the definition of the class $\Psi_{1,2}$ follows that
\begin{gather*}
\int_{0}^{t}\int_B\mid y(e^{\psi(s,y)}-1)\mid ds\nu(dy)=\int_{[0,t]\times B\cap\underline{A}}\mid y(e^{\psi(s,y)}-1)\mid ds\nu(dy)\\[1ex]
+\int_{[0,t]\times B\cap\hat{A}}\mid y(e^{\psi(s,y)}-1)\mid ds\nu(dy)\\[1ex]
\leq \left(\int_{0}^{t}\int_{B}\mid y\mid^2 ds\nu(dy)\right)^{\frac{1}{2}}\left(\int_{\underline{A}}\mid e^{\psi(s,y)}-1\mid^2 ds\nu(dy)\right)^{\frac{1}{2}}
+\int_{\hat{A}}\mid e^{\psi(s,y)}-1\mid ds\nu(dy)<+\infty,
\end{gather*}
where $B:=\{y: \mid y\mid\leq1\}$ and thus $\tilde{a}_t$ is actually well defined. The compensating measure $\nu_Q(ds,dy)$  satisfies 
\begin{gather}\label{war na miare Levyego przy Q}
\int_{0}^{t}\int_{\mathbb{R}}(\mid y\mid^2\wedge \ 1)\nu_Q(ds,dy)<+\infty,
\end{gather}
 because
\begin{gather*}
\int_{0}^{t}\int_{\mathbb{R}}(\mid y\mid^2\wedge \ 1)e^{\psi(s,y)} \ ds\nu(dy)<+\infty \quad \Longleftrightarrow \quad \int_{0}^{t}\int_{\mathbb{R}}(\mid y\mid^2\wedge \ 1)\mid e^{\psi(s,y)}-1\mid \ ds\nu(dy)<+\infty,
\end{gather*}
and using a similar decomposition with the sets $\underline{A}$ and $\hat{A}$ as above we obtain
\begin{align*}
\int_{0}^{t}\int_{\mathbb{R}}(\mid y\mid^2\wedge \ 1)\mid e^{\psi(s,y)}-1\mid ds\nu(dy)&\leq \int_{0}^{t}\int_{\mathbb{R}}(\mid y\mid^2\wedge \ 1)ds\nu(dy)\\[1ex]
&+\int_{\hat{A}}\mid e^{\psi(s,y)}-1\mid ds\nu(dy)<+\infty,
\end{align*}
hence all the terms in \eqref{Z przy mierze Q} are well defined. Actually \eqref{Z przy mierze Q} follows immediately from the L\'evy-It\^o decomposition \eqref{tw Levy Ito decomposition} by adding and subtracting  the terms $\int \phi ds$ and  $\int\int_{B}ye^{\psi}ds\nu(dy)$. 

It follows that under $Q$ the process $Z$ is a L\'evy process only if $\phi$ is a deterministic constant and $\psi$ is a deterministic function independent on time, that is
$\phi(\omega,t)=\phi$, $\psi(\omega,t,y)=\psi(y)$. This is a very particular situation and it follows that, in general, $Z$ is not a L\'evy process under $Q$ any more. In particular, the measure $Q$ changes stochastic properties of the jumps of $Z$ because the new compensating measure $\nu_Q$ is random and time dependent. Hence $\pi$ is no longer a Poisson random measure under $Q$. 

It is clear that under $Q$ the small jumps are square summable and there are only finite  number of big jumps on $[0,T^\ast]$ because $Q\sim P$. However, the condition \eqref{war na miare Levyego przy Q} does not imply that the corresponding expectations are finite just like it was under the physical measure $P$. Since the integral in \eqref{war na miare  Levyego przy Q} is continuous in $t$, it follows from \eqref{war na miare  Levyego przy Q} that there exists a localizing sequence of stopping times $\{\tau_n, n=1,2,...\}$ such that 
$$
E^Q\Big[\int_{0}^{\tau_n}\int_{\mathbb{R}}(\mid y\mid^2\wedge \ 1)\nu_Q(ds,dy)\Big]<+\infty, \quad n=1,2,...,
$$
which implies that
\begin{gather*}
E^Q\Big[\sum_{s\in[0,\tau_n]}\mid\triangle Z_s\mid^2\mathbf{1}_{\{\mid\triangle Z_s\mid\leq1\}}\Big]<+\infty, \quad E^Q\Big[\sum_{s\in[0,\tau_n]}\mathbf{1}_{\{\mid\triangle Z_s\mid>1\}}\Big]<+\infty, \quad n=1,2,... \ .
\end{gather*}
Moreover, for any set $A\in \mathbb{R}$ with $0\notin \bar{A}$ holds
\begin{gather}\label{skonczonosc miary kompensujacej na A przy Q}
\nu_Q([0,t],A)=\int_{0}^{t}\int_{A}e^{\psi(s,y)}ds\nu(dy)<+\infty,
\end{gather}
and using similar arguments as above one can show that the process
$$
\tilde{\pi}_{Q}(t,A), \qquad t\geq0,
$$
is a $Q$-local martingale and not a $Q$-martingale in general. The property \eqref{skonczonosc miary kompensujacej na A przy Q} follows from the estimation
$$
\int_{0}^{t}\int_{A}\mid e^{\psi(s,y)}-1\mid ds\nu(dy)\leq t\nu(A)+\int_{\hat{A}}\mid e^{\psi(s,y)}-1\mid ds\nu(dy)<+\infty.
$$

\section{Martingale representation under equivalent measures}\label{Martingale representation under equivalent measures}

As explained in Section \ref{section Martingale representation and characterisation of equivalent measures} the process $Z$, which is a L\'evy process under $P$, is not a L\'evy process under an equivalent measure $Q$ any more.  Its jump measure is not a Poisson measure and hence  Theorem \ref{tw Kunity} can not be applied for $Q$-local martingales. Our aim is to formulate an analogue result to Theorem \ref{tw Kunity} and to this aim also to construct the integral over the compensated jump measure of $Z$ under $Q$. A comprehensive exposition of this part of the theory is missing in the literature.  

\subsection{Integration over the compensated jump measure under $Q$}\label{section Construction of the integral}

Let $(\phi,\psi)$ be a generating pair of a measure $Q\sim P$.  In view of Theorem 
\ref{tw Girsanova} the jump measure $\pi(dt,dy)$ of $Z$ has a new compensating measure under $Q$ of the form $\nu_Q(dt,dy)=e^{\psi(t,y)}dt\nu(dy)$. Consequently, $\tilde{\pi}_Q(ds,dy)=\pi(ds,dy)-e^{\psi(s,y)}ds\nu(dy)$ is a compensated jump measure of $Z$ under $Q$. Our aim now is to construct the stochastic integral 
\begin{gather}\label{calka po dpiq}
\int_{0}^{t}\int_{\mathbb{R}} g(s,y)\tilde{\pi}_Q(ds,dy), \qquad t\in[0,T^\ast],
\end{gather}
for $g:[0,T^\ast]\times \mathbb{R}\longrightarrow \mathbb{R}$. We will start from simple processes and then extend 
the construction to a wider class of integrands. First notice that
$$
E^Q[\pi(t,A)]=E^Q[\nu_Q(t,A)]=E^Q\Big[\int_{0}^{t}\int_{A}e^{\psi(s,y)}ds\nu(dy)\Big], \quad t\geq0,
$$
may be infinite even if the set $A$ is separated from zero. For that reason we introduce an additional restriction on the separated from zero subsets of $\mathbb{R}$, that is
 \begin{gather}\label{warunek na A przy Q}
E^Q[\nu_Q([0,T^\ast]\times A)]<+\infty.
\end{gather}
If \eqref{warunek na A przy Q} holds then the process $\tilde{\pi}_{Q}(t,A)=\pi(t,A)-\nu_Q(t,A)$ is a $Q$-martingale although its increments are not independent. In fact $\tilde{\pi}_Q(t,A)$ is then a $Q$-square integrable martingale and its properties are formulated below. 
 
\begin{lem}\label{lem wlasnosci piq}
 Let $A,B\subseteq \mathbb{R}$ be such that for both \eqref{warunek na A przy Q} holds. Then the processes 
 $\tilde{\pi}_Q(t,A)$, $\tilde{\pi}_Q(t,B)$ are square integrable martingales under $Q$ on $[0,T^\ast]$ and
 their quadratic covariation is given by
\begin{gather}\label{wahanie kwadratowe przy Q}
  \mbox{\boldmath $\langle$}\tilde{\pi}_Q(t,A),\tilde{\pi}_Q(t,B)\mbox{\boldmath $\rangle$}=\nu_Q([0,t]\times A\cap B), \quad t\in[0,T^\ast].
\end{gather}
In particular, the processes
\begin{align*}
(\tilde{\pi}_Q(t,A))^2-\nu_Q([0,t]\times A); \qquad 
\tilde{\pi}_Q(t,A)\cdot\tilde{\pi}_Q(t,B)-\nu_Q([0,t],A\cap B), \quad t\in[0,T^\ast]
\end{align*}
are $Q$-martingales and if $A\cap B=\emptyset$ then 
$$
E^Q[\tilde{\pi}_Q(t,A)\cdot\tilde{\pi}_Q(t,B)]=0, \quad t\in[0,T^\ast].
$$
\end{lem}
{\bf Proof:} First let us notice that $\tilde{\pi}_Q(t,A)$,  $\tilde{\pi}_Q(t,B)$ are $Q$-locally square integrable martingales.  Indeed, the process $\pi(t,A)$ has jumps of size $1$ and $\nu_Q([0,t]\times A)$ is continuous, so both are locally bounded and thus $\tilde{\pi}_Q(t,A)$ is locally bounded, hence $Q$-locally square integrable. It follows that the process $\mbox{\boldmath $\langle$}\tilde{\pi}_Q(t,A),\tilde{\pi}_Q(t,B)\mbox{\boldmath $\rangle$}$ is well defined. Application of the It\^o product formula, see Theorem 4.4.13 in \cite{Appl}, yields
\begin{align*}
 \tilde{\pi}_Q(t,A)\cdot\tilde{\pi}_Q(t,B)&=\int_{0}^{t}\tilde{\pi}_Q(s-,A)d\tilde{\pi}_Q(s,B)+\int_{0}^{t}\tilde{\pi}_Q(s-,B)d\tilde{\pi}_Q(s,A)\\[1ex]
 &+\sum_{s\in[0,t]}\triangle\tilde{\pi}_Q(s,A)\cdot \triangle\tilde{\pi}_Q(s,B), \qquad t\in[0,T^\ast].
\end{align*}
The first two integrals on the right side are $Q$-local martingales as stochastic integrals of locally bounded processes with respect to martingales. Since both processes $\tilde{\pi}_Q(t,A),\tilde{\pi}_Q(t,B)$ have jumps of size $1$ we have 
$$
\sum_{s\in[0,t]}\triangle\tilde{\pi}_Q(t,A)\cdot \triangle\tilde{\pi}_Q(t,B)=\pi([0,t]\times A\cap B), \quad t\in[0,T^\ast].
$$
Compensating the last term we obtain
$$
\sum_{s\in[0,t]}\triangle\tilde{\pi}_Q(t,A)\cdot \triangle\tilde{\pi}_Q(t,B)=\tilde{\pi}_Q([0,t]\times A\cap B)+\nu_Q([0,t]\times A\cap B),\quad t\in[0,T^\ast].
$$
Finally, the process 
$$
\tilde{\pi}_Q(t,A)\cdot\tilde{\pi}_Q(t,B)-\nu_Q([0,t]\times A\cap B), \qquad t\in[0,T^\ast],
$$
is a $Q$-local martingale which gives \eqref{wahanie kwadratowe przy Q}. Further it follows from the estimation
$$
E^Q[\tilde{\pi}_Q(T^\ast,A)^2]=E^Q[\nu_Q([0,T^\ast]\times A)]<+\infty,
$$
that  $\tilde{\pi}_Q(t,A)$ is in fact a $Q$-square integrable martingale.
\hfill$\square$
\vskip2mm

In the first step one constructs the integral \eqref{calka po dpiq} for a simple process having the representation
\begin{gather}\label{g simple Q}
 g(s,y)=g(0,y)\mathbf{1}_{\{s=0\}}+\sum_{i=0}^{n-1}\left(\sum_{j=1}^{m_i}g_{ij}\mathbf{1}_{(t_i,t_{i+1}]}(s)\mathbf{1}_{A_{ij}}\right), \qquad s\in[0,T^\ast], \ y\in U,
\end{gather}
where $0=t_0<t_1<...<t_n=T^\ast$ is a partition of $[0,T^\ast]$ and $A_{i,j}$ is a family of subsets of $\mathbb{R}$ separated from zero such that 
\begin{gather}\label{warunek na Aij przy Q}
E ^Q[\nu_Q([0,T^\ast]\times A_{i,j})]<+\infty.
\end{gather}
The set of simple processes, denoted by $\mathcal{S}^Q$, is similar to $\mathcal{S}$ defined in Section \ref{section Construction of the integral}. The difference lies
in the condition \eqref{warunek na Aij przy Q} imposed on the sets $A_{ij}$ and this requirement is related to the different form of the compensating measure under $Q$. Actually, under $P$, the analogue of \eqref{warunek na Aij przy Q} holds automatically if only $\{A_{ij}\}$ are separated from zero. It turns out that for $g\in\mathcal{S}^Q$ the stochastic integral
$$
I^Q(g)_t=\int_{0}^{t}\int_{U}g(s,y)\tilde{\pi}_{Q}(ds,dy):=\sum_{i=0}^{n}\sum_{j=1}^{m_i}g_{ij}\tilde{\pi}_{Q}((t_i\wedge t,t_{i+1}\wedge t]\times A_{ij}), \qquad t\in[0,T^\ast].
$$
is a $Q$-square integrable martingale. This can be proved having in hand Proposition \ref{prop o wwo under Q} below which is a counterpart of Proposition \ref{prop o wwo under P} and describes properties of the $Q$-compensated jump measure. Its proof is directly based on Lemma \ref{lem wlasnosci piq} and is left to the reader. 

\begin{prop}\label{prop o wwo under Q}
 For the sets $A,B\in U$ satisfying \eqref{warunek na A przy Q} and $0\leq s<t\leq T^\ast]$ hold  
 \begin{align*}
  &E^{Q}[\tilde{\pi}_{Q}^2\left((s,t]\times A\right)\mid\mathcal{F}_{s}]=E^{Q}[\nu_Q\left((s,t]\times A\right)\mid\mathcal{F}_{s}],\\[1ex]
  &E^Q[\tilde{\pi}_{Q}\left((s,t]\times A\right)\cdot \tilde{\pi}_{Q}\left((s,t]\times B\right)\mid \mathcal{F}_s]=0, \quad \text{if} \ A\cap B=\emptyset, \\[1ex]
  &E^Q[\tilde{\pi}_{Q}\left((s,t]\times A\right)\cdot \tilde{\pi}_{Q}\left((u,v]\times B\right)\mid \mathcal{F}_u]=0, \quad \text{for} \ t\leq u<v\leq T^\ast.
 \end{align*}
\end{prop}
Then mimicking the proof of Proposition \ref{prop izometria dla prostych przy P} one can prove the following.
\begin{prop}
 For a process $g\in\mathcal{S}^Q$ the integral $I^Q(g)$ is a $Q$-square integrable martingale and
 \begin{gather}\label{wzor izometryczny dla prostych przy Q}
   E^Q\left[\big\vert\int_{0}^{t}\int_{\mathbb{R}}g(s,y)\tilde{\pi}_{Q}(ds,dy)\big\vert^2\right]=E^Q\left[\int_{0}^{t}\int_{\mathbb{R}}\mid g(s,y)\mid^2\nu_Q(ds,dy)\right], \quad t\in[0,T^\ast].
 \end{gather}
\end{prop}
To extend the definition of the integral on larger class of integrands we use the same arguments as under the measure $P$.  
If a predictable process $g$ satisfies
$$
E^{Q}\Big[\int_{0}^{T^\ast}\int_{\mathbb{R}}\mid g(s,y)\mid^2\nu_Q(ds,dy)\Big]<+\infty.
$$
then $I^Q(g)$ is defined by the approximating arguments. It is a $Q$-square integrable martingale and \eqref{wzor izometryczny dla prostych przy Q} holds. By $\Psi^Q_2$ we denote the class of all predictable processes satisfying
$$
\Psi^Q_2: \qquad \int_{0}^{T^\ast}\int_{\mathbb{R}}\mid g(s,y)\mid^2\nu_Q(ds,dy)<+\infty, \qquad Q-a.s.
$$
For $g\in\Psi^Q_2$ the integral $I^Q(g)$ is a $Q$-locally square integrable martingale. If $g$ satisfies the condition
\begin{gather}\label{EQ skonczona}
E^Q\left[\int_{0}^{T^\ast}\int_{\mathbb{R}}\mid g(s,y)\mid\nu_Q(ds,dy)\right]<+\infty.
\end{gather}
then one defines 
\begin{gather}\label{def skompensowanej calki przy Q}
\int_{0}^{t}\int_{\mathbb{R}}g(s,y)\tilde{\pi}_Q(ds,dy):=\int_{0}^{t}\int_{\mathbb{R}}g(s,y){\pi}(ds,dy)-\int_{0}^{t}\int_{\mathbb{R}}g(s,y)\nu_Q(ds,dy), \qquad t\in[0,T^\ast],
\end{gather}
which is a $Q$-martingale. If $g\in\Psi^Q_1$, where
$$
\Psi^Q_1: \qquad \int_{0}^{T^\ast}\int_{\mathbb{R}}\mid g(s,y)\mid\nu_Q(ds,dy)<+\infty, \qquad Q-a.s.,
$$
then \eqref{EQ skonczona} holds locally and thus the process
$$
\int_{0}^{t}\int_{\mathbb{R}}g(s,y)\tilde{\pi}_Q(ds,dy), \quad t\in[0,T^\ast],
$$
is a $Q$-local martingale.

Finally, for the representation of $Q$- local martingales we need a class $\Psi^Q_{1,2}$ defined by
$$
g\in\Psi^Q_{1,2} \qquad \Longleftrightarrow \qquad g\mathbf{1}_{\{\mid g\mid\leq 1\}}\in\Psi^Q_2 \quad \text{and}\quad g\mathbf{1}_{\{\mid g\mid> 1\}}\in\Psi^Q_1.
$$
For each $g\in\Psi^Q_{1,2}$ the integral is defined by the decomposition
$$
\int_{0}^{t}\int_{\mathbb{R}}g(s,y)\tilde{\pi}_{Q}(ds,dy):=\int_{0}^{t}\int_{\mathbb{R}}g\mathbf{1}_{\{\mid g\mid\leq 1\}}\tilde{\pi}_{Q}(ds,dy)
+\int_{0}^{t}\int_{\mathbb{R}}g\mathbf{1}_{\{\mid g\mid> 1\}}\tilde{\pi}_{Q}(ds,dy),
$$
and is a $Q$-local martingale. It is clear that $g\in\Psi^Q_{1,2}$ if and only if
$$
\int_{0}^{T^\ast}\int_{U}(\mid g(s,y)\mid^2\wedge \mid g(s,y)\mid)\nu_Q(ds,dy)<+\infty, \quad Q-a.s..
$$

\subsection{Martingale representation under $Q$}

Using the class of integrands described in Section \ref{section Construction of the integral} we can decompose any $Q$-local martingale to the integral form.

\begin{tw}\label{tw o reprezentacji przy mierze Q}
Let $Q$ be a measure equivalent to $P$ with a generating pair $(\phi,\psi)$,  $\phi\in\Phi$ and	
\begin{gather}\label{e do psi -1 w Psi1}
e^{\psi}-1\in\Psi_{1}.
\end{gather}
Any $Q$-local martingale $M_{t},t\in[0,T^\ast],$ admits a representation of the form
\begin{gather}\label{reprezentacja Kunity przy Q}
M_t=M_0+\int_{0}^{t}\tilde{\phi}_M(s) d\tilde{W}(s)+\int_{0}^{t}\int_{\mathbb{R}}\tilde{\psi}_{M}(s,y)\tilde{\pi}_{Q}(ds,dy), \quad t\in[0,T^\ast]
\end{gather}
with $\tilde{\phi}_M\in\Phi$ and $\tilde{\psi}_M\in\Psi^Q_{1,2}$. Moreover, the pair $(\tilde{\phi}_M,\tilde{\psi}_M)$ is unique i.e., if
$(\tilde{\phi}_M^{\prime},\tilde{\psi}_M^{\prime})$ satisfies \eqref{reprezentacja Kunity przy Q} then
\begin{gather*}
\tilde{\phi}_M=\tilde{\phi}_M^{\prime} \quad dQ\times dt- \ \text{a.s.}  \quad
\text{and} \quad \tilde{\psi}_M=\tilde{\psi}_M^{'}   \quad
dQ\times d\nu_Q- \ \text{a.s.}.
\end{gather*}
\end{tw}

In the proof we will use the classical result, for its the proof see, for instance, Proposition 3.8 in \cite{JacShir}.

\begin{lem}\label{lem P and Q local martingales} Let $Q$ be equivalent to $P$ and have the density process
$\rho_t:=\frac{dQ}{dP}\mid_{\mathcal{F}_t}, t\in[0,T^\ast]$. Then the process $M(t)$ is a $Q$-local martingale if and only if $M(t)\rho(t)$ is a $P$-local martingale.
\end{lem}

\vskip2mm
\noindent
{\bf Proof: }\emph{[of Theorem \ref{tw o reprezentacji przy mierze Q}]} We consider the case with no Wiener part, that is the density process
has the form
\begin{gather}\label{rho}
d\rho(t)=\rho(t-)\int_{\mathbb{R}}(e^{\psi(t,y)}-1)\tilde{\pi}(dt,dy), \quad \rho(0)=1, \quad t\in[0,T^\ast].
\end{gather}
The passage to the general case does not cause serious difficulties.
In view of Lemma \ref{lem P and Q local martingales} the process $\rho_t M_t, t\in[0,T^\ast]$ is a $P$-local martingale and by Theorem \ref{tw Kunity} admits the representation
\begin{gather}\label{reprezentacja Mrho}
\rho_t M_t=\rho_0M_0+\int_{0}^{t}\int_{\mathbb{R}}\psi_M(s,y)\tilde{\pi}(ds,dy), \quad t\in[0,T^\ast],
\end{gather}
for some $\psi_M\in\Psi_{1,2}$. From the It\^o formula and \eqref{rho} follows 
\begin{align}\nonumber
\frac{1}{\rho_t}&=1-\int_{0}^{t}\frac{1}{\rho_{s-}^2}d\rho_s+\sum_{s\in[0,t]}\left\{\frac{1}{\rho_s}-\frac{1}{\rho_{s-}}+\frac{1}{\rho^2_{s-}}\triangle\rho_s\right\}\\[1ex]\nonumber
&=1-\int_{0}^{t}\int_{\mathbb{R}}\frac{1}{\rho_{s-}}(e^{\psi(s,y)}-1)\tilde{\pi}(ds,dy)\\[1ex]\label{1 przez rho}
&+\int_{0}^{t}\int_{\mathbb{R}}\frac{1}{\rho_{s-}}\left(e^{-\psi(s,y)}+e^{\psi(s,y)}-2\right)\pi(ds,dy), \quad t\in[0,T^\ast].
\end{align}

\noindent
 Application of the It\^o product formula together with \eqref{reprezentacja Mrho} and \eqref{1 przez rho} yields
\begin{align*}
M_t&=(M_t\rho_t)\cdot\frac{1}{\rho_t}=M_0+\int_{0}^{t}(M\rho)_{s-}d\big(\frac{1}{\rho_s}\big)+\int_{0}^{t}\frac{1}{\rho_{s-}}d(M\rho)_s+\big[M\rho,\frac{1}{\rho}\big]_s\\[1ex]
&=M_0+\int_{0}^{t}\int_{\mathbb{R}}M_{s-}\Big((e^{-\psi}+e^{\psi}-2)\pi(ds,dy)-(e^{\psi}-1)\tilde{\pi}(ds,dy)\Big)\\[1ex]
&+\int_{0}^{t}\int_{\mathbb{R}}\frac{1}{\rho_{s-}}\Big(\psi_M(e^{-\psi}-1)\pi(ds,dy)+\psi_M\tilde{\pi}(ds,dy)\Big), \quad t\in[0,T^\ast].
\end{align*}
Now, we use the fact that $e^{\psi(t,y)}dt\nu(dy)$ is a compensating measure of $\pi(dt,dy)$ under $Q$ and rearrange the terms above. This gives
$$
M_t=M_0+\int_{0}^{t}\int_{\mathbb{R}}M_{s-}e^{-\psi(s,y)}(1-e^{\psi(s,y)})\tilde{\pi}_Q(ds,dy)+\int_{0}^{t}\int_{\mathbb{R}}
\frac{1}{\rho_{s-}}\psi_M(s,y)e^{-\psi(s,y)}\tilde{\pi}_Q(ds,dy).
$$
The proof is completed by showing that the integrals above are actually well defined, that is that the process
$$
\tilde{\psi}_M(s,y):=M_{s-}e^{-\psi(s,y)}(1-e^{\psi(s,y)})+\frac{1}{\rho_{s-}}e^{-\psi(s,y)}\psi_M(s,y), \quad t\in[0,T^\ast], y\in U,
$$
belongs to $\Psi_{1,2}^Q$. Since the processes $M$ and $\frac{1}{\rho}$ are locally bounded, we will prove that
\begin{gather}\label{pierwszy warunek calkowy}
e^{-\psi}(1-e^{\psi})\in\Psi_{1,2}^Q,
\end{gather}
and
\begin{gather}\label{drugi warunek calkowy}
e^{-\psi}\psi_M\in\Psi_{1,2}^Q.
\end{gather}
The condition \eqref{pierwszy warunek calkowy} follows from the estimation
$$
\Big(\mid e^{-\psi}(1-e^{\psi})\mid^2\wedge \mid e^{-\psi}(1-e^{\psi})\mid \Big)e^{\psi}=e^{-\psi}\mid e^{\psi}-1\mid^2\wedge\mid e^{\psi}-1\mid\leq\mid e^{\psi}-1\mid
$$
and \eqref{e do psi -1 w Psi1}. The condition \eqref{drugi warunek calkowy} has the form
\begin{gather}\label{drugi warunek calkowy przeformulowany}
\int_{0}^{T^\ast}\int_{\mathbb{R}}H(s,y)ds\nu(dy)<+\infty,\quad \text{with} \
H(s,y):=\mid\psi_M(s,y)\mid^2e^{-\psi(s,y)}\wedge \mid\psi_M(s,y)\mid, 
\end{gather}
and in view of \eqref{e do psi -1 w Psi1} we need to prove \eqref{drugi warunek calkowy przeformulowany} in the case when $\psi\leq0$ only. Let us consider the following subsets of $[0,T^\ast]\times \mathbb{R}$
\begin{gather*}
A:=\{\psi\leq 0\},\quad
B:=\{\mid \psi_M\mid^2e^{-\psi}\leq\mid\psi_M\mid\},\quad
C:=\{e^{\psi}\leq\frac{1}{2}\}.
\end{gather*}
From \eqref{e do psi -1 w Psi1} follows that 
$$
\int_{0}^{T^\ast}\int_{\mathbb{R}}\mathbf{1}_{A\cap B}(1-e^{\psi(s,y)})ds\nu(dy)<+\infty,
$$
and thus $A\cap C$ is a set of finite $dt\nu(dy)$ measure. The four estimations below
\begin{enumerate}[a)]
\item
\begin{align*}
\int_{0}^{T^\ast}\int_{\mathbb{R}}&\mathbf{1}_{A\cap B\cap C}H(s,y)ds\nu(dy)	
=\int_{0}^{T^\ast}\int_{\mathbb{R}}\mathbf{1}_{A\cap B\cap C}\mid\psi_M(s,y)\mid^2e^{-\psi(s,y)}ds\nu(dy),\\[1ex]
&\leq\int_{0}^{T^\ast}\int_{\mathbb{R}}\mathbf{1}_{A\cap B\cap C}\mid\psi_M(s,y)\mid ds\nu(dy)
\leq\frac{1}{2}\int_{0}^{T^\ast}\int_{\mathbb{R}}\mathbf{1}_{A\cap B\cap C}ds\nu(dy)<+\infty,
\end{align*}
because $A\cap B\cap C$ is of finite measure,
\item 
\begin{align*}
\int_{0}^{T^\ast}\int_{\mathbb{R}}&\mathbf{1}_{A\cap B\cap C^c}H(s,y)ds\nu(dy)\leq
\int_{0}^{T^\ast}\int_{\mathbb{R}}\mathbf{1}_{A\cap B\cap C^c}\Big(2\mid\psi_M(s,y)\mid^2\wedge \mid\psi_M(s,y)\mid\Big)ds\nu(dy)\\[1ex]
&\leq 2\int_{0}^{T^\ast}\int_{\mathbb{R}}\mathbf{1}_{A\cap B\cap C^c}\Big(\mid\psi_M(s,y)\mid^2\wedge \mid\psi_M(s,y)\mid\Big)ds\nu(dy)<+\infty,
\end{align*}
because $\psi_M\in\Psi_{1,2}$,
\item 
\begin{align*}
\int_{0}^{T^\ast}\int_{\mathbb{R}}&\mathbf{1}_{A\cap B^c\cap C^c}H(s,y)ds\nu(dy)\leq
\int_{0}^{T^\ast}\int_{\mathbb{R}}\mathbf{1}_{A\cap B^c\cap C^c}\Big(2\mid\psi_M(s,y)\mid^2\wedge \mid\psi_M(s,y)\mid\Big)ds\nu(dy)\\[1ex]
&\leq 2\int_{0}^{T^\ast}\int_{\mathbb{R}}\mathbf{1}_{A\cap B^c\cap C^c}\Big(\mid\psi_M(s,y)\mid^2\wedge \mid\psi_M(s,y)\mid\Big)ds\nu(dy)<+\infty,
\end{align*}
\item 
\begin{align*}
\int_{0}^{T^\ast}\int_{\mathbb{R}}&\mathbf{1}_{A\cap B^c\cap C}H(s,y)ds\nu(dy)=
\int_{0}^{T^\ast}\int_{\mathbb{R}}\mathbf{1}_{A\cap B^c\cap C\cap\{\mid\psi_M\mid\leq1\}}\mid\psi_M(s,y)\mid ds\nu(dy)\\[1ex]
&+\int_{0}^{T^\ast}\int_{\mathbb{R}}\mathbf{1}_{A\cap B^c\cap C\cap\{\mid\psi_M\mid>1\}}\mid\psi_M(s,y)\mid ds\nu(dy)\\[1ex]
&\leq\int_{0}^{T^\ast}\int_{\mathbb{R}}\mathbf{1}_{A\cap B^c\cap C\cap\{\mid\psi_M\mid\leq1\}}ds\nu(dy)\\[1ex]
&+\int_{0}^{T^\ast}\int_{\mathbb{R}}\mathbf{1}_{A\cap B^c\cap C\cap\{\mid\psi_M\mid>1\}}\Big(\mid\psi_M(s,y)\mid^2\wedge\mid\psi_M(s,y)\mid\Big)ds\nu(dy)<+\infty,
\end{align*}
\end{enumerate}
imply \eqref{drugi warunek calkowy przeformulowany}. The uniqueness is equivalent to the implication
$$
M_t:=0=\int_{0}^{t}\int_{\mathbb{R}}\tilde{\psi}_M(s,y)\tilde{\pi}_Q(dy,ds), \quad t\in[0,T^\ast], \quad \tilde{\psi}_M\in\Psi_{1,2}^Q \quad \Longrightarrow \quad \tilde{\psi}_M\equiv0.
$$
From the It\^o formula follows
\begin{align*}
0=M_t\rho_t&=\int_{0}^{t}\int_{\mathbb{R}}M_{s-}\rho_{s-}(e^{\psi(s,y)}-1)d\tilde{\pi}(ds,dy)+\int_{0}^{t}\int_{U}\rho_{s-}\tilde{\psi}_M(s,y)\tilde{\pi}_Q(dy,ds)\\[1ex]
&+\int_{0}^{t}\int_{\mathbb{R}}
\rho_{s-}\tilde{\psi}_M(s,y)(e^{\psi(s,y)}-1)\pi(ds,dy)\\[1ex]
&=\int_{0}^{t}\int_{\mathbb{R}}\rho_{s-}\tilde{\psi}_M(s,y)e^{\psi(s,y)}\tilde{\pi}(ds,dy), \quad t\in[0,T^\ast],
\end{align*}
and the uniqueness of the integral representation under $P$ implies that $\tilde{\psi}_M\equiv0$.
\hfill$\square$
\vskip2mm

The decomposition \eqref{reprezentacja Kunity przy Q} has been already formulated, without the uniqueness property, in \cite{Kun}, see Theorem 2.3. The proof in \cite{Kun} is based on another arguments and is, however, sketchy. In particular, it is not clear in \cite{Kun} which 
processes can be integrated over the compensated jump measure under $Q$.

\section{Incompleteness of the bond market}\label{Incompleteness of the bond market}
We will examine the problem of completeness of the bond market. Our main result is Theorem \ref{tw_o_niezupelnosci} showing the market incompleteness in the case when the  L\'evy measure has a density function. This result generalizes Theorem 4.12 in \cite{BarZab IJTAF} where the model was specified under the martingale measure, that is $P$ was a martingale measure.

\subsection{The model}
The market under consideration consists of bonds with maturities forming a set $[0,T^\ast]$ with $T^\ast<+\infty$. For any $T\in[0,T^\ast]$ the price of the $T$-bond is defined by 
\begin{gather}\label{P postac}
P(t,T):=e^{-\int_{t}^{T}f(t,u)du}, \quad t\in[0,T^\ast],
\end{gather}
where $f(\cdot,\cdot)$ stands for a forward rate. The time evolution of the forward rate is defined for each $T\in[0,T^\ast]$ separately by
\begin{gather}\label{forward rate dynamics}
f(t,T)=f(0,T)+\int_{0}^{t}\alpha(s,T)ds+\int_{0}^{t}\sigma(s,T)dZ(s), \quad t\in[0,T^\ast],
\end{gather}
where $Z$ is a L\'evy process on $(\Omega,\mathcal{F}, (\mathcal{F}_{t})_{t\in[0,T^\ast]}, P)$. We adopt the model assumptions from \cite{EbJacRaib}, that is the coefficients in \eqref{forward rate dynamics} satisfy
\begin{align}\label{A1}
\alpha(t,T)=0, \  \sigma(t,T)=0 \quad \text{for} \ t\in[T,T^\ast],\\[2ex]\label{A2}
(\omega,t,T)\longrightarrow \alpha(\omega,t,T), \sigma(\omega,t,T) \quad \text{are} \ \mathcal{P}\otimes\mathcal{B}([0,T^\ast])-\text{measurable},\\[2ex]\label{A3}
\sup_{t,T\in[0,T^\ast]}\mid\alpha(t,T)\mid<+\infty,  \sup_{t,T\in[0,T^\ast]}\mid\sigma(t,T)\mid<+\infty.
\end{align}
In \eqref{A2} $\mathcal{P}$ stands for the predictable $\sigma$-field on $\Omega\times[0,T^\ast]$ and $\mathcal{B}([0,T^\ast])$ for the Borel $\sigma$-field
on $[0,T^\ast]$. In view of \eqref{A3} the fields $(t,T)\longrightarrow \alpha(t,T), \sigma(t,T)$ are assumed to be bounded, but the bound may depend on $\omega$. Under \eqref{A3} both integrals in \eqref{forward rate dynamics} are well defined. Finally, \eqref{A1} allows to define the bond prices for time points exceeding their maturities. To see this define the short rate process by $r(t):=f(t,t), t\in[0,T^\ast]$. Then, by \eqref{A1} and \eqref{forward rate dynamics}, we have
$$
f(t,T)=f(0,T)+\int_{0}^{T}\alpha(s,T)ds+\int_{0}^{T}\sigma(s,T)dZ(s)=f(T,T), \quad t\in[T,T^\ast],
$$
and consequently
\begin{align*}
P(t,T)=e^{-\int_{t}^{T}f(t,s)ds}=e^{-\int_{t}^{T}f(s,s)ds}=e^{\int_{T}^{t}r(s)ds}=e^{\int_{T}^{t}r(s)ds}, \qquad
 t\in[T,T^\ast].
\end{align*}
The latter condition means that the nominal value of the bond is automatically transferred at maturity on the savings account and stays there till $T^\ast$. Further, it follows from \eqref{A1} that the discounted bond prices
$$
\hat{P}(t,T):=e^{-\int_{0}^{t}r(s)ds}P(t,T), \quad t,T\in[0,T^\ast],
$$
can be represented in the form
\begin{gather}\label{P discounted}
\hat{P}(t,T):=e^{-\int_{0}^{T}f(t,s)ds},\quad t,T\in[0,T^\ast].
\end{gather}
The issue of prime importance is the absence of arbitrage in the model defined by \eqref{P postac} - \eqref{forward rate dynamics}.
The concept of arbitrage, in the sense of \cite{DelSch1} and \cite{DelSch2}, amounts to the existence of a measure $Q$ which is equivalent to $P$ and
such that the discounted bond prices are $Q$-local martingales. Each such a measure is called a martingale measure.
The following result, which is a consequence of Theorem 3.1 in \cite{EbJacRaib} and Theorem 3.1 in \cite{Jakubowski-Zabczyk}, specifies the relation between the generating pair $(\phi,\psi)$ of the martingale measure and the model coefficients.

\begin{tw}\label{tw HJM} 
Assume that \eqref{A1}-\eqref{A3} are satisfied. Let $Q\sim P$ be a measure with a generating pair  $(\phi, \psi)$ such that $\phi\in\Phi$, $e^{\psi}-1\in\Psi_{1,2}$. Denote
$$
A(t,T):=\int_{t}^{T}\alpha(t,v)dv, \qquad \Sigma(t,T):=\int_{t}^{T}\sigma(t,v)dv,\qquad t,T\in[0,T^\ast].
$$

\begin{enumerate}[a)]
\item If the processes $\hat{P}(\cdot,T), T\in [0,T^\ast]$ given by \eqref{P discounted} are $Q$-local martingales then 
\begin{gather}\label{HJM war konieczny 1}
\int_{0}^{T^\ast}\int_{\{\mid y\mid\leq 1\}}\mid e^{\psi(s,y)}-1\mid ds \ \nu(dy)<+\infty, \quad a.s.,
\end{gather}
and
\begin{gather}\label{HJM war konieczny 2}
\int_{0}^{T^\ast}\int_{\{\mid y\mid>1\}}e^{-\Sigma(s,T) y}\cdot e^{\psi(s,y)}ds \ \nu(dy)<+\infty,
\end{gather}
for each $T$ almost $\omega$-surely.
\item If \eqref{HJM war konieczny 1} and \eqref{HJM war konieczny 2} are satisfied then $\hat{P}(\cdot,T), T\in[0,T^\ast]$ are $Q$-local martingales if and only if
\begin{align}\label{warunek HJM physical measure general}\nonumber
A(s,T)&=-\Sigma(s,T)a+\frac{1}{2}q\Sigma(s,T)^2-q\phi(s)\Sigma(s,T)\\[1ex]
&+\int_{\mathbb{R}}\big(e^{\psi(s,y)}(e^{-\Sigma(s,T)y}-1)+\mathbf{1}_{\{\mid y\mid\leq 1\}}\Sigma(s,T)y\big)\nu(dy),
\end{align} 
for each $T\in[0,T^\ast]$ almost all $s$  almost $\omega$-surely. 
\end{enumerate}
\end{tw}

Let us comment on the theorem above. The conditions \eqref{HJM war konieczny 1} and \eqref{HJM war konieczny 2} narrow the class of generating pairs of  martingale measures. Actually it follows from \eqref{HJM war konieczny 1} and the fact that $e^\psi-1\in\Psi_{1,2}$ that 
\begin{gather}\label{e do psi -1 w klasie Psi1}
e^{\psi}-1\in\Psi_1.
\end{gather}
The condition \eqref{HJM war konieczny 2} is a generalization of the exponential moment conditions obtained in \cite{Jakubowski-Zabczyk} for the case when $\phi=0,\psi=0$, that is when the model is specified directly under the martingale measure. Notice that the right side of \eqref{warunek HJM physical measure general} involves the volatility of the forward rate, the characteristics of $Z$ and the generating pair of $Q$ while the left side depends on the drift of the forward rate only. Differentiation of \eqref{warunek HJM physical measure general}  in $T$ gives a direct formula for $\alpha(t,T)$, which generalizes the famous Heath-Jarrow-Morton drift condition 
$$
\alpha(t,T)=\sigma(t,T)\int_{t}^{T}\sigma(t,v)dv,\quad t,T\in[0,T^\ast],
$$
introduced in \cite{HeatJarrMor} in the case when $Z$ was a Wiener process.
 
 Let $\mathcal{Q}$ be a set of all martingale measures and $Q\in\mathcal{Q}$ have a generating pair $(\phi,\psi)$. Recall, that under $Q$ the process
$\tilde{W}$ given by \eqref{tilde W} is a Wiener process under, $\nu_Q(dt,dy)$ given by  
\eqref{nu_Q} is a compensating measure of $\pi(dt,dy)$ and $\tilde{\pi}_Q(dt,dy)=\pi(dt,dy)-\nu_Q(dt,dy)$ is a compensated jump measure of $Z$.
 The use of \eqref{P discounted}, \eqref{forward rate dynamics} together with the It\^o formula provides the dynamics of $\hat{P}(\cdot,T), T\in[0,T^\ast]$ under $Q$. It has the form
\begin{align}\label{wzor reprezentacja P^ under Q}\nonumber
\hat{P}(t,T)&=\hat{P}(t,0)-\int_{0}^{t}\hat{P}(s-,T)\Sigma(s,T)d\tilde{W}(s)\\[1ex]
&+\int_{0}^{t}\int_{\mathbb{R}}\hat{P}(s-,T)\left[e^{-\Sigma(s,T)y}-1\right]\tilde{\pi}_Q(ds,dy), \quad t,T\in[0,T^\ast].
\end{align}

\subsection{Portfolios}
The concept of a bond portfolio generalizes the finite dimensional setting involved in the stock market description. From 
\eqref{P postac} follows that for any $t\in[0,T^\ast]$ the function
$$
T\rightarrow P(t,T), \quad T\in[0,T^\ast],
$$
is continuous and hence $P_t:=P(t,\cdot)$ can be treated as an element of a Banach space $B$ of bounded functions on $[0,T^\ast]$
with norm
$$
\ | h|_B:=\sup_{z\in[0,T^\ast]}\mid h(z)\mid. 
$$
A trading strategy $\varphi$ will be a $B^\ast$-valued process, where $B^\ast$ stands for the dual of $B$.
The corresponding wealth process is defined by
$$
X^\varphi_t=\langle \varphi_t,P_t \rangle_B, \qquad t\in[0,T^\ast],
$$ 
where $\langle \varphi,P \rangle_B$ is the value of the functional $\varphi\in B^\ast$ on the element $P\in B$. Consequently, the discounted wealth process is given by
$$
\hat{X}^\varphi_t=\langle \varphi_t,\hat{P}_t \rangle_B, \qquad t\in[0,T^\ast].
$$ 
In the class of self-financing strategies the changes of the portfolio value arise from the fluctuations of the bond price process which means that
$\hat{X}^{\varphi}$ admits the following representation then 
\begin{gather}\label{X hat dynamcs}
\hat{X}^{\varphi}_t=\hat{X}^{\varphi}_0+\int_{0}^{t}\langle\varphi_s,d\hat{P}_s\rangle_B, \quad t\in[0,T^\ast],
\end{gather}
where the latter integral is to be precisely defined. Taking into account \eqref{wzor reprezentacja P^ under Q} we have
\begin{gather}\label{portfel przy mierze martyngalowej}
\hat{X}^{\varphi}_t=\hat{X}^\varphi_0-\int_{0}^{t}\langle\varphi_s,\hat{P}_{s-}\Sigma_s\rangle_Bd\tilde{W}_s
+\int_{0}^{t}\int_{\mathbb{R}}\langle\varphi_s,\hat{P}_{s-}(e^{-\Sigma_s y}-1)\rangle_B\tilde{\pi}_Q(ds,dy), \quad t\in[0,T^\ast],
\end{gather}
 where $\Sigma_s:=\Sigma(s,\cdot)\in B$.
This leads to the following definition of an admissible strategy.

\begin{def}\label{def strategi dopuszczaych} Let $Q\in\mathcal{Q}$ be a martingale measure with a generating pair $(\phi,\psi)$, $\phi\in\Phi$, $e^\psi-1\in\Psi_{1}$. A $B^\ast$-valued strategy $\varphi$ is admissible if
\begin{enumerate}[a)] 
 \item $\langle\varphi_s,\hat{P}_{s-}\Sigma_s\rangle_B\in\Phi$ and $\langle\varphi_s,\hat{P}_{s-}(e^{-\Sigma_s y}-1)\rangle_B\in\Psi^Q_{1,2}$, that is 
 \begin{gather}\label{pierwszy warunek na dopuszczalnosc}
  \int_{0}^{T^\ast}\mid\langle\varphi_s,\hat{P}_{s-}\Sigma_s\rangle_B\mid^2 ds<+\infty,
 \end{gather}
 and 
 \begin{gather}\label{drugi warunek na dopuszczalnosc}
  \int_{0}^{T^\ast}\int_{\mathbb{R}}\Big(
  \mid\langle\varphi_s,\hat{P}_{s-}(e^{-\Sigma_s y}-1)\rangle_B\mid^2\wedge\mid\langle\varphi_s,\hat{P}_{s-}(e^{-\Sigma_s y}-1)\rangle_B\mid\Big) e^{\psi(s,y)}ds\nu(dy)<+\infty,
 \end{gather}
 \item the wealth process, which is given by \eqref{portfel przy mierze martyngalowej}, is a $Q$-martingale, 
\end{enumerate}
The class of all admissible trading strategies will be denoted by $\mathcal{A}(Q)$.
\end{def}
The definition of admissible strategies depends on the choice of martingale measure. This feature of the model is caused by the presence of jumps. Indeed, although \eqref{pierwszy warunek na dopuszczalnosc} is not measure dependent, \eqref{drugi warunek na dopuszczalnosc} is. Since the arbitrage free model may admit many martingale measures the definition above may seem confusing because it involves only one fixed measure.
The relevance of the definition can be, however,  justified by the use of economical arguments admitting the model framework where 
the prices of financial contracts are given by expectations under a so called {\it pricing measure} which is chosen from the set of all martingale measures. The methods of choosing the pricing measure will be not discussed here, we mention only that the model framework with a unique pricing measure is often used in practice and also in purely theoretical consideration. Alternatively we could also take into account trading strategies from the set 
$$
\mathcal{A}:=\bigcap_{Q\in\mathcal{Q}}\mathcal{A}_{Q}.
$$
The problem is, however,  that $\mathcal{A}$ can be significantly smaller than $\mathcal{A}(Q)$ and consequently the set of investing possibilities would become poor. Also for that reason we fix only one martingale measure and consider admissible strategies related to that measure.

\subsection{Incompleteness}\label{Incompleteness} 
Let us start from the definition of the market completeness.

\begin{dfn} The bond market defined by \eqref{P postac}-\eqref{forward rate dynamics} is complete if for each bounded random variable $X$ there exists $\varphi\in\mathcal{A}(Q)$ such that 
\begin{gather}\label{warunek na replikacje}
X=\hat{X}^{\varphi}_{T^\ast}.
\end{gather}
A strategy satisfying \eqref{warunek na replikacje} is called a replicating strategy for $X$. The market is incomplete if it is not complete.
\end{dfn}
Let $X$ be a bounded random variable. In view of Theorem \ref{tw o reprezentacji przy mierze Q} the associated martingale $M_t:=E^Q[X\mid\mathcal{F}_t], t\in[0,T^\ast]$ admits the integral representation 
$$
X=E^Q[X]+\int_{0}^{T^\ast}f_X(s)d\tilde{W}(s)+\int_{0}^{T^\ast}g_X(s,y)\tilde{\pi}_Q(ds,dy),
$$
for some $f_X\in\Phi$ and $g_X\in\Psi_{1,2}^Q$. By \eqref{portfel przy mierze martyngalowej} the discounted wealth process of an admissible strategy at $T^\ast$ is given by
\begin{gather*}
\hat{X}^{\varphi}_{T^\ast}=\hat{X}^\varphi_0-\int_{0}^{T^\ast}\langle\varphi_s,\hat{P}_{s-}\Sigma_s\rangle_Bd\tilde{W}_s
+\int_{0}^{T^\ast}\int_{\mathbb{R}}\langle\varphi_s,\hat{P}_{s-}(e^{-\Sigma_s y}-1)\rangle_B\tilde{\pi}_Q(ds,dy).
\end{gather*}
Since $\hat{X}^\varphi$ is a $Q$-martingale, the strategy $\varphi$ replicates $X$ if and only if the following conditions are satisfied
\begin{align}\label{warunek na replikacje 1}
\hat{X}^\varphi_0&=E^Q[X],\\[1ex]\label{warunek na replikacje 2}
-\langle\varphi_s,\hat{P}_{s-}\Sigma_s\rangle_B&=f_X(s), \quad dt-a.s.,\\[1ex]\label{warunek na replikacje 3}
\langle\varphi_s,\hat{P}_{s-}(e^{-\Sigma_s y}-1)\rangle_B&=g_X(s,y), \quad dQ\times d\nu_Q-a.s..
\end{align}  
If $\varphi$ replicates $X$ then \eqref{warunek na replikacje 1} defines the replication cost. The solution $\varphi$ to \eqref{warunek na replikacje 2} can be searched in the class of functionals which are point evaluations, that is for any maturity $T\in[0,T^\ast]$ consider
$$
\langle\delta_{T},h\rangle_B:=h(T), \quad h\in B.
$$
Under trivial non-degeneracy conditions there exists a solution $c=c(s)$ of the equation
$$
-c(s)\hat{P}(s-,T)\Sigma(s,T)=f_X(s), \quad s\in[0,T^\ast] 
$$
and hence $\varphi(s):=c(s)\delta_{T}$ solves \eqref{warunek na replikacje 2}. What requires a deeper analysis is the condition \eqref{warunek na replikacje 3} which involves the jumps of the driving process $Z$. Although \eqref{warunek na replikacje 3} must hold $dQ\times d\nu_Q-a.s.$, we will
use in the sequel the concept of a concentration point of the original  L\'evy measure $\nu$ of $Z$. The precise definition, which has been introduced in \cite{BarZab IJTAF}, is as follows.

\begin{dfn}\label{def concetr. point}
A point $y_0\in \mathbb{R}$ is a concentration point of the L\'evy measure
$\nu$ if there exists a sequence $\{\varepsilon_n\}_{n=1}^{\infty}$
s.t. $\varepsilon_n\searrow 0$ satisfying
\begin{gather}\label{war na punkt koncetracji}
\nu\Big\{B(y_0,\varepsilon_n)\backslash
B(y_0,\varepsilon_{n+1})\Big\}>0 \quad \forall \ n=1,2,...,
\end{gather}
where $B(y_0,\varepsilon)=\{y\in \mathbb{R}: 
|y-y_0|\leq\varepsilon\}$.
\end{dfn}

The definition above captures a great majority of L\'evy processes used in the financial modelling. Indeed, each L\'evy process for which its L\'evy measure has a density function has also a concentration point.
Our aim now is to prove that if the L\'evy measure has a concentration point then \eqref{warunek na replikacje 3} has no solution for some $g_X$ and consequently the bond market model is incomplete. 

\begin{tw}\label{tw_o_niezupelnosci}
Consider the bond market model \eqref{P postac}-\eqref{forward rate dynamics} with coefficients satisfying \eqref{A1}-\eqref{A3}.
If the L\'evy measure $\nu$ of $Z$ has a concentration point
$y_0\neq 0$
then the market is incomplete.
\end{tw}

In the proof we will use two auxiliary results formulated below. The first is an extension of the moment problem solution, see Theorem 2 in Section 5 of Yosida \cite{Yos} or Lemma 4.5 in \cite{BarZab IJTAF}.
\begin{lem}\label{lemat o momentach}
Let $E $ be a normed linear space and $A$ an arbitrary set.
Let $g:A\longrightarrow \mathbb{R}$ and
$h:A\longrightarrow E $. Then there exists $e^\ast\in
E^\ast$ such that
\begin{gather}\label{warunek na funkcjonal}
g(a)=<e^\ast,h(a)>_{E}, \quad \forall a\in A,
\end{gather}
if and only if
\begin{gather}\nonumber
\exists \ \gamma>0 \quad \forall \ n\in\mathbb{N} \quad \forall \
\{\beta_i\}_{i=1}^{n}, \ \beta_i\in\mathbb{R} \quad \forall \
\{a_i\}_{i=1}^{n}, \ a_i\in A  \quad  holds\\[2ex]\label{warunek z momentow}
\Big|\sum_{i=1}^{n}\beta_i g(a_i)\Big|\leq\gamma \
\Big\vert\sum_{i=1}^{n}\beta_i h(a_i){\Big\vert}_{E}.
\end{gather}
\end{lem}

The second result follows from the Fubini theorem and will simplify examining of the condition \eqref{warunek na replikacje 3}.

\begin{prop}\label{prop o repr. produktowej}
Let $(E_1,\mathcal{E}_1,\mu_1)$, $(E_2,\mathcal{E}_2,\mu_2)$ be
measurable spaces with sigma-finite measures $\mu_1, \mu_2$ and
$(E_1\times
E_2,\mathcal{E}_1\times\mathcal{E}_2,\mu_1\times\mu_2)$ be their
product space. If two measurable functions $f_1:E_1\times
E_2\longrightarrow \mathbb{R}$, $f_2:E_1\times E_2\longrightarrow
\mathbb{R}$ satisfy the condition
\begin{gather}\label{rownosc funkcji}
f_1=f_2, \qquad  d\mu_1\times d\mu_2- \text{a.s.},
\end{gather}
then there exists a set $\hat{E}_1\in\mathcal{E}_1$ such that
\begin{gather}\label{1war na E1}
\hat{E}_1 \quad {\text is \ of \ full \ \mu_1 \
measure}\\\label{2war na E1} \forall x\in\hat{E}_1  \quad  {\text
the \ set}\quad \{y: f_1(x,y)=f_2(x,y)\}\quad {\text is \ of \ full
\ \mu_2 \ measure}.
\end{gather}
\end{prop}

\vskip2mm
\noindent
{\bf Proof:} \emph{[of Theorem \ref{tw_o_niezupelnosci}]} We will construct a bounded random variable $X$ such that there is no admissible strategy solving \eqref{warunek na replikacje 3}.

Let $\{\varepsilon_n\}_{n=1}^{\infty}$ be a sequence satisfying
\eqref{war na punkt koncetracji} and define an auxiliary deterministic function
$g$ by the formula
\[
g(y)=
\begin{cases}
| y|\wedge \ 1 \quad &\text{for} \quad
y\in\{B(y_0,\varepsilon_{2k+1})\backslash
B(y_0,\varepsilon_{2k+2})\} \quad k=0,1,...,
\\[2ex]
-(| y|\wedge \ 1) \quad &\text{for} \quad
y\in\{B(x_0,\varepsilon_{2k})\backslash B(x_0,\varepsilon_{2k+1})\}
\quad k=1,2,...,\\[2ex]
| y| \wedge \ 1 \quad &\text{for} \quad
y\in(-\infty,y_0-\varepsilon_1)\cup(y_0+\varepsilon_1)\cup\{y_0\}.
\end{cases}
\]
Assume that for some $\varphi\in\mathcal{A}(Q)$ holds
\begin{gather}\label{rownanie z g}
\langle\varphi_t,\hat{P}_{t-}(e^{-\Sigma_t y}-1)\rangle_B=g(y),
\end{gather}
$dQ\times d\nu_Q-a.s.$. Since the measures $dQ\times \nu_Q(dt,dy)$ and $dP\times dt \times d\nu$  are equivalent, the equality 
\eqref{rownanie z g} holds $dP\times dt \times d\nu$ a.s.. Fix $(\omega,t)\in\Omega\times[0,T^\ast]$ and assume that \eqref{rownanie z g} holds $\nu$-a.s.. From Proposition \ref{prop o repr. produktowej} follows that then there exists a set $A_{\nu}(\omega,t)$ of a
full $\nu$-measure s.t. \eqref{rownanie z g} is satisfied for each $y\in A_{\nu}(\omega,t)$. Due to
Lemma \ref{lemat o momentach} there exists
$\gamma=\gamma(\omega,t)>0$ such that
\begin{gather}\label{war momentow w dowodzie}\nonumber
\forall \ n\in\mathbb{N} \quad \forall \ \{\beta_i\}_{i=1}^{n}, \
\beta_i\in\mathbb{R} \quad \forall \ \{y_i\}_{i=1}^{n}, \ y_i\in
A_{\nu}(\omega,t)\\[2ex]\label{war_tw}
\Big|\sum_{i=1}^{n}\beta_ig(y_i)\Big|\leq\gamma\Big\vert
\sum_{i=1}^{n}\beta_i \hat{P}_{t-}(e^{-\Sigma_t y_i}-1) \Big\vert_{B}.
\end{gather}
Let us notice that due to \eqref{war na punkt koncetracji} we have
\begin{gather*}
\nu\Big\{A_{\nu}(\omega,t)\cap \big\{B(x_0,\varepsilon_n)\backslash
B(x_0,\varepsilon_{n+1})\big\}\Big\}>0
\end{gather*}
so we can choose a sequence $\{a_k\}_{k=1}^{\infty}$ s.t.
\begin{gather*}
a_k\in A_{\nu}(\omega,t)\cap \big\{B(y_0,\varepsilon_k)\backslash
B(y_0,\varepsilon_{k+1})\big\} \quad  \forall \ k=1,2,... .
\end{gather*}
\noindent Let us examine condition \eqref{war_tw} with $n=2$,
$\beta_1=1, \beta_2=-1$ and $y_1=a_{2k+1}$, $y_2=a_{2k+2}$ for
$k=0,1,...$. Then the left side of \eqref{war_tw} has a
form
\begin{gather*}
\frac{1}{\gamma}\Big|\beta_1g(a_{2k+1})+\beta_2g(a_{2k+2})\Big|=\frac{1}{\gamma}
\Big((|a_{2k+1}| \wedge \ 1)+(|a_{2k+2}| \mid \wedge \
1)\Big)
\end{gather*}
and thus satisfies
\begin{gather*}
\lim_{k\longrightarrow\infty}\frac{1}{\gamma}\Big|\beta_1g(a_{2k+1})+\beta_2g(a_{2k+2})\Big|=\frac{2(|
y_0|\wedge \ 1)}{\gamma}\neq 0.
\end{gather*}
\noindent Now let us estimate the right side of \eqref{war_tw}.
\begin{align*}
\Big\vert\hat{P}_{t-}(e^{-\Sigma_t a_{2k+1}}-1)&-\hat{P}_{t-}(e^{-\Sigma_t a_{2k+2}}-1)\Big\vert_{B}\\[2ex]
&= \sup_{T\in[0,T^\ast]}\Big|\hat{P}(t-,T)(e^{-\Sigma(t,T)a_{2k+1}}-1)-\hat{P}(t-,T)(e^{-\Sigma(t,T)a_{2k+1}}-1)\Big|\\[2ex]
&\leq \sup_{T\in [0,T^\ast]}|\hat{P}(t-,T)| \ \sup_{T\in [0,T^\ast]}
\Big|e^{\Sigma(t,T)a_{2k+1}}-e^{-\Sigma(t,T)a_{2k+2}}\Big|.
\end{align*}
The first supremum is clearly finite. To deal with the second supremum let us notice
that for sufficiently large $k$ the points $a_{2k+1},a_{2k+2}$ lie
in $B(y_0,\delta);\delta>0$ and thus we have
\begin{align}\label{nier na pochodna}\nonumber
\sup_{T\in [0,T^\ast]}
&\Big|e^{-\Sigma(t,T)a_{2k+1}}-e^{-\Sigma(t,T)a_{2k+2}}\Big|\\[1ex]\nonumber
&\leq\sup_{T\in [0,T^\ast]}\sup_{y\in
B(y_0,\delta)}\Big|De^{-\Sigma(t,T)y}\Big|\cdot|
a_{2k+1}-a_{2k+2}|\\[2ex]
&\leq\sup_{T\in [0,T^\ast]}\sup_{y\in B(y_0,\delta)}\left\{e^{|\Sigma(t,T)|\cdot |y|}
\cdot |\Sigma(t,T)|\right\}\cdot|a_{2k+1}-a_{2k+2}|,
\end{align}
which clearly tends to zero. Thus condition \eqref{war_tw} is not
satisfied for any $(\omega,t)\in \Omega\times[0,T^\ast]$ and thus
\eqref{rownanie z g} does not hold $\nu - a.s.$ for
any $(\omega,t)\in \Omega\times[0,T^\ast]$. As a consequence of
Proposition \ref{prop o repr. produktowej} the equation \eqref{rownanie z g} does not have a solution.

\noindent Now, with the use of the function $g$, we construct a
bounded random variable $X$ which can not be replicated. First let us notice that for a martingale measure $Q$ with a generating pair $(\phi,\psi)$ we have
$$
\int_{0}^{T^\ast}\int_{\mathbb{R}}(\mid g(y)\mid^2\wedge\mid g(y)\mid)e^{\psi(s,y)}\nu(dy)ds
\leq \int_{0}^{T^\ast}\int_{\mathbb{R}}(\mid y\mid^2\wedge 1)e^{\psi(s,y)}\nu(dy)ds<+\infty,
$$
see \eqref{war na miare Levyego przy Q}, which means that $g\in\Psi^Q_{1,2}$. Let $\tau_k$ be a stopping time defined by
\begin{gather*}
\tau_k=\inf\{t:
\Big\arrowvert\int_{0}^{t}\int_{\mathbb{R}}g(y)\tilde{\pi}_{Q}(ds,dy)\Big\arrowvert\geq
k\}\wedge T^{\ast},
\end{gather*}
and choose a number $k_0$ s.t. the set
$\{(\omega,\tau_{k_0}(\omega));
\omega\in\Omega\}\subseteq\Omega\times[0,T^\ast]$ is of positive
$dP\times dt$- measure. Then the process
$g_X(s,y):=g(y)\mathbf{1}_{(0,\tau_{k_0}]}(s)$ is predictable, bounded and belongs to $\Psi^Q_{1,2}$. The process $\int_{0}^{\cdot}\int_{\mathbb{R}}g_X(s,y)\tilde{\pi}_{Q}(ds,dy)$ is bounded because $|\Delta\int_{0}^{\cdot}\int_{\mathbb{R}}g_X(x,y)\tilde{\pi}_Q(ds,dy)|\leq1$ holds
and thus is a $Q$-martingale as a bounded $Q$-local martingale. Finally let us define 
\begin{gather}\label{zmienna niereplikowalna ograniczona}
X:=\int_{0}^{T^{\ast}}\int_{\mathbb{R}}g_X(s,y)\tilde{\pi}_Q(ds,dy).
\end{gather} 
Then \eqref{warunek na replikacje 3} has no solution in the class of admissible strategies and hence $X$ can not be replicated.
\hfill$\square$

\vskip2mm

Let us notice that the issue of uniqueness of the martingale measure does not affect the thesis of Theorem \ref{tw_o_niezupelnosci}, that is the market is incomplete even if the martingale measure $Q$ is unique. This shows a significant difference comparing to the stock market with a finite number of assets  where the equivalence between the uniqueness of the martingale measure  and the market completeness is one of the basic properties.

\end{document}